\documentclass[11pt]{llncs}

\usepackage{epsfig}
\usepackage{algorithm}
\usepackage{algorithmic}
\usepackage{subfigure}
\usepackage{amssymb}
\usepackage{amsmath}
\usepackage{graphicx}
\usepackage{paralist}
\usepackage{xspace}
\usepackage{tabularx}
\usepackage{latexsym}
\usepackage{times}
\usepackage{color}

\graphicspath{{figures/}} 

\newcommand{\belong}{{\sc Partitioned $2$-page Book Embedding}\xspace}
\newcommand{\be}{{\sc P2BE}\xspace}

\newcommand{\skel}{skel}
\newcommand{\pert}{pert}
\newcommand{\remove}[1]{}

\begin{document}


\title{Implementing a Partitioned $2$-page Book Embedding Testing
Algorithm\thanks{This work was partially supported by the ESF project
10-EuroGIGA-OP-003 GraDR ``Graph Drawings and Representations'', by the
MIUR of Italy, under project AlgoDEEP, prot. 2008TFBWL4, and by the
italian inter-university computing Consortium CASPUR.}}

\author{
  Patrizio Angelini$^1$
  \and
  Marco Di Bartolomeo$^{1,2}$
  \and
  Giuseppe Di Battista$^1$
}

\institute{
  Dip.~di Informatica e Automazione, Roma Tre University, Italy \and
  Italian Inter-University Computing Consortium CASPUR, Italy
\email{\{angelini,gdb\}@dia.uniroma3.it, m.dibartolomeo@caspur.it}
}

\maketitle

\begin{abstract}
In a book embedding the vertices of a graph are placed on the ``spine'' of
a ``book'' and the edges are assigned to ``pages'' so that edges on the
same page do not cross.
In the \belong problem egdes are partitioned
into two sets $E_1$ and $E_2$, the pages are two, the edges of $E_1$
are assigned to page $1$, and the edges of $E_2$ are assigned to
page $2$. The problem consists of checking if an ordering of the vertices
exists along the spine so that the edges of each page do not cross.
Hong and Nagamochi~\cite{hn-tpbesg-09tr} give an interesting and complex
linear time algorithm for tackling \belong
based on SPQR-trees.
We show an efficient implementation of this algorithm and show its
effectiveness by performing a number of
experimental tests.
Because of the relationships~\cite{hn-tpbesg-09tr} between \belong and
clustered planarity we yield as a side effect an
implementation of a clustered planarity testing where the graph has exactly two clusters.
\end{abstract}

\section{Introduction}

In a \emph{book embedding}~\cite{o-btvg-73} of a graph the vertices are
placed on the ``spine'' of a ``book'' and
the edges are assigned to ``pages'' so that edges on the same page do not cross. 
A rich body of literature witnesses the interest of the scientific community for book
embeddings. See,
e.g., \cite{bs-pnpg-84,y-epgfp-89}.

Several constrained variations of book embeddings have been studied. 
In~\cite{w-dcbe-02} the problem is tackled when in each page the number of
edges incident to a vertex is bounded. 
In~\cite{ffr-opnupda-12} the graph is directed upward planar and the order
of the vertices on the spine must be consistent with the orientation of the
edges. 
Hong and Nagamochi~\cite{hn-tpbesg-09tr} provide a linear time algorithm for
a problem called \emph{\sc Partitioned $2$-page Book Embedding} (\emph{\sc
P2BE}). In
the \be problem the egdes of an input graph $G(V,E_1,E_2)$ are partitioned
into two sets $E_1$ and $E_2$, the pages are just two, the edges of $E_1$
are assigned to page $1$, and the edges of $E_2$ are assigned to
page $2$. The problem consists of checking if an ordering of the vertices
exists along the spine so that the edges of each page do not cross.

In~\cite{hn-tpbesg-09tr} the \be problem is characterized in terms of
the existence of an embedding of $G$ allowing to build a variation of
the dual graph containing a particular Eulerian tour.
The existence of such an embedding is tested exploiting
SPQR-trees~\cite{dt-opt-96} for biconnected components and
BC-trees for connected ones.

In this paper we discuss an implementation of the algorithm
in~\cite{hn-tpbesg-09tr}. To efficiently implement the algorithm
we faced the following problems: (i) One of the key steps of the algorithm
requires the enumeration and the analysis of all the permutations of a set
of objects. Even if the cardinality of the set is bounded by a constant
this may lead to very long execution times. We restated that step of the
algorithm avoiding such enumerations. (ii) Some steps of the
algorithm are described in~\cite{hn-tpbesg-09tr} at a high abstraction
level. We found how to efficiently
implement all of them. (iii) The algorithm builds several
embeddings that are tested for the required properties only at the
end of the computation. Our implementation considers only one embedding
that is greedily built to have the properties. We performed experiments
over a large set of suitably randomized graphs. The experiments show quite
reasonable linear execution times.

The algorithm in~\cite{hn-tpbesg-09tr} is interesting in itself, since book
embedding problems are ubiquitous in Graph Drawing. However, it is even more
appealing because it yields~\cite{hn-tpbesg-09tr} almost immediately a
linear time algorithm for the following special case of \emph{clustered
planarity} testing. A planar graph $G(V_1,V_2,E)$ whose vertices are
partitioned into two sets (clusters) $V_1$ and $V_2$ is given. Is it
possible to find a planar drawing for $G$ such
that:
\begin{inparaenum}[(i)]
\item each of $V_1$ and $V_2$ is drawn inside a simple region,
\item the two regions are disjoint, and
\item each edge of $E$ crosses the boundary of a region at most once?
\end{inparaenum}
Using the terminology of Clustered Planarity, this is a clustered
planarity testing for a flat clustered graphs with exactly two clusters.
References on clustered planarity can be found, e.g.,
in~\cite{fce-pcg-95,CorteseB05}.
Hence, we yield, as a side effect, an
implementation of such special case of clustered planarity testing.
An alternative algorithm for the same clustered planarity problem has been
proposed
in~\cite{Biedl:1998:DPP,b-dppIIItcep-98}.

The paper is organized as follows.
In Section~\ref{se:preliminaries} we give preliminaries.
In Section~\ref{se:book-embedding} we outline the algorithm.
Section~\ref{se:spqr} discusses how to search an embedding with the desired features and
Section~\ref{se:nodes} gives further implementation details on the search.
In Section~\ref{se:experiments} we describe our experiments.
Section~\ref{se:conclusions} gives concluding remarks.

%

\section{Preliminaries}\label{se:preliminaries}

In this section we give preliminary definitions that will be used in the paper.

\subsection{Planarity}

A \emph{planar drawing} of a graph is a mapping of each vertex to a
distinct point of the plane and of each edge to a simple Jordan curve
connecting its endpoints such that the curves representing the edges
do not cross but, possibly, at common endpoints. A graph is \emph{planar}
if it admits a planar drawing. Two drawings of a graph are
\emph{equivalent} if they determine the same circular ordering around each
vertex. An \emph{embedding} is an equivalence class of
drawings. A planar drawing partitions the plane into topologically
connected regions, called  {\em faces}. The unbounded face is the
\emph{outer face}.

\subsection{Connectivity and SPQR-trees}

A graph is \emph{connected} if every two vertices are joined by a
path. A graph $G$ is \emph{biconnected} (\emph{triconnected}) if
removing any vertex (any two vertices) leaves $G$ connected.

To handle the decomposition of a biconnected graph into its
triconnected components, we use \emph{SPQR-trees}
(see~\cite{dt-omtc-96,dt-opt-96,gm-lti-00}).

A graph is \emph{st-biconnectible} if adding edge $(s,t)$ to it yields a
biconnected graph. Let $G$ be an st-biconnectible graph. A \emph{separation
pair} of $G$ is a pair of vertices whose removal disconnects the graph. A
\emph{split pair} of $G$ is either a separation pair or a pair of adjacent
vertices. A \emph{maximal split component} of $G$ with respect to a split
pair $\{u, v\}$ (or, simply, a maximal split component of $\{u, v\}$) is
either an edge $(u, v)$ or a maximal subgraph $G'$ of $G$ such that $G'$
contains $u$ and $v$, and $\{u, v\}$ is not a split pair of $G'$. A vertex
$w \neq u,v$ belongs to exactly one maximal split component of $\{u, v\}$.
We call \emph{split component} of $\{u, v\}$ the union of any number of
maximal split components of $\{u, v\}$.

We assume consider SPQR-trees that are rooted at one edge of the graph,
called the \emph{reference edge}.

The rooted SPQR-tree $\mathcal{T}$ of a biconnected graph $G$, with respect
to a reference edge $e$, describes a recursive decomposition of $G$ induced
by its split pairs. The nodes of $\mathcal{T}$ are of four types: S, P, Q,
and R. Their connections are called \emph{arcs}, in order to distinguish
them from the edges of $G$.

Each node $\mu$ of $\mathcal{T}$ has an associated st-biconnectible
multigraph, called the \emph{skeleton} of $\mu$ and denoted by
\skel($\mu$). Skeleton \skel($\mu$) shows how the children of $\mu$,
represented by ``virtual edges'', are arranged into $\mu$. The virtual edge
in \skel($\mu$) associated with a child node $\nu$, is called the
\emph{virtual edge of $\nu$ in \skel($\mu$)}.

For each virtual edge $e_i$ of \skel($\mu$), recursively replace $e_i$ with
the skeleton \skel($\mu_i$) of its corresponding child $\mu_i$. The
subgraph of $G$ that is obtained in this way is the \emph{pertinent graph}
of $\mu$ and is denoted by \pert($\mu$).

Given a biconnected graph $G$ and a reference edge $e=(u',v')$, tree
$\mathcal{T}$ is recursively defined as follows. At each step, a split
component $G^*$, a pair of vertices $\{u,v\}$, and a node $\nu$ in
$\mathcal{T}$ are given. A node $\mu$ corresponding to $G^*$ is introduced
in $\mathcal{T}$ and attached to its parent $\nu$. Vertices $u$ and $v$ are
the \emph{poles} of $\mu$ and denoted by $u(\mu)$ and $v(\mu)$,
respectively. The decomposition possibly recurs on some split components of
$G^*$. At the beginning of the decomposition $G^* = G - \{e\}$,
$\{u,v\}=\{u',v'\}$, and $\nu$ is a Q-node corresponding to $e$.

\begin{description}

\item[\textbf{Base Case:}] If $G^*$ consists of exactly one edge between
$u$ and $v$, then $\mu$ is a Q-node whose skeleton is $G^*$ itself.

\item[\textbf{Parallel Case:}] If $G^*$ is composed of at least two maximal
split components $G_1, \dots, G_{k}$ ($k \geq 2$) of $G$ with respect to
$\{u,v\}$, then $\mu$ is a P-node. Graph \skel($\mu$) consists of $k$
parallel virtual edges between $u$ and $v$, denoted by $e_1, \dots, e_{k}$
and corresponding to $G_1, \dots, G_{k}$, respectively. The decomposition
recurs on $G_1, \dots, G_{k}$, with $\{u,v\}$ as pair of vertices for every
graph, and with $\mu$ as parent node.

\item[\textbf{Series Case:}] If $G^*$ is composed of exactly one maximal
split component of $G$ with respect to $\{u,v\}$ and if $G^*$ has
cutvertices $c_1, \dots, c_{k-1}$ ($k \geq 2$), appearing in this order on
a path from $u$ to $v$, then $\mu$ is an S-node. Graph \skel($\mu$) is the
path $e_1, \dots, e_k$, where virtual edge $e_i$ connects $c_{i-1}$ with
$c_i$ ($i = 2, \dots ,k-1$), $e_1$ connects $u$ with $c_1$, and $e_k$
connects $c_{k-1}$ with $v$. The decomposition recurs on the split
components corresponding to each of $e_1, e_2,\dots, e_{k-1}, e_{k}$ with
$\mu$ as parent node, and with $\{u,c_1\}, \{c_1,c_2\},$ $\dots,$
$\{c_{k-2},c_{k-1}\}, \{c_{k-1},v\}$ as pair of vertices, respectively.

\item[\textbf{Rigid Case:}] If none of the above cases applies, the purpose
of the decomposition step is that of partitioning $G^*$ into the minimum
number of split components and recurring on each of them. We need some
further definition. Given a maximal split component $G'$ of a split pair
$\{s,t\}$ of $G^*$, a vertex $w \in G'$ \emph{properly belongs} to $G'$ if
$w \neq s, t$. Given a split pair $\{s,t\}$ of $G^*$, a maximal split
component $G'$ of $\{s,t\}$ is \emph{internal} if neither $u$ nor $v$ (the
poles of $G^*$) properly belongs to~$G'$, \emph{external} otherwise. A
\emph{maximal split pair} $\{s,t\}$ of $G^*$ is a split pair of $G^*$ that
is not contained into an internal maximal split component of any other
split pair $\{s',t'\}$ of $G^*$. Let $\{u_1,v_1\}, \dots, \{u_k,v_k\}$ be
the maximal split pairs of $G^*$ ($k \geq 1$) and, for $i = 1, \dots, k$,
let $G_i$ be the union of all the internal maximal split components of
$\{u_i,v_i\}$. Observe that each vertex of $G^*$ either properly belongs to
exactly one $G_i$ or belongs to some maximal split pair $\{u_i,v_i\}$. Node
$\mu$ is an R-node. Graph \skel($\mu$) is the graph obtained from $G^*$ by
replacing each subgraph $G_i$ with the virtual edge $e_i$ between $u_i$ and
$v_i$. The decomposition recurs on each $G_i$ with $\mu$ as parent node and
with $\{u_i,v_i\}$ as pair of vertices.
\end{description}

For each node $\mu$ of $\mathcal{T}$, the construction of \skel($\mu$) is
completed by adding a virtual edge $(u,v)$ representing the rest of the
graph.

The SPQR-tree $\mathcal{T}$ of a graph $G$ with $n$ vertices and $m$ edges
has $m$ Q-nodes and $O(n)$ S-, P-, and R-nodes. Also, the total number of
vertices of the skeletons stored at the nodes of $\mathcal{T}$ is $O(n)$.
Finally, SPQR-trees can be constructed and handled efficiently. Namely,
given a biconnected planar graph $G$, the SPQR-tree $\mathcal{T}$ of $G$
can be computed in linear time~\cite{dt-omtc-96,dt-opt-96,gm-lti-00}.

\subsection{Book Embedding}

A \emph{book embedding} of a graph $G=(V,E)$ consists of a total
ordering of the vertices in $V$ and of an assignment of the
edges in $E$ to \emph{pages}, in such a way that no two
edges $(a,b)$ and $(c,d)$ are assigned to the same page if $a\prec c
\prec b \prec d$. A \emph{$k$-page book embedding} is a book embedding
using $k$ pages. A \emph{partitioned $k$-page book embedding} is a
$k$-page book embedding in which the assignment of edges to the pages
is part of the input. In the special case when $k=2$, we call the problem
\belong (\be). Hence, an instance of the \belong problem is just a graph
$G(V,E_1,E_2)$, whose edges are partitioned into two sets $E_1$ and
$E_2$, the pages are just two, and the edges of $E_1$ are pre-assigned to
page $1$ and the edges of $E_2$ are pre-assigned to page $2$. We say that
the edges of $E_1$ (of $E_2$) are \emph{red} (\emph{blue}) edges. 

\subsection{Eulerian Tour}

Let $G$ be a directed planar embedded graph. A directed cycle of $G$ is a
\emph{Eulerian tour} if it traverses each edge exactly once. Consider a
vertex $v$ of $G$ and let $(v_1,v)$, $(v,v_2)$, $(v,v_3)$, and $(v_4,v)$ be
four edges incident to $v$ appearing in this order around $v$ in the given
embedding. If a Eulerian tour contains egdes $(v_1,v)$, $(v,v_3)$,
$(v_4,v)$, and $(v,v_2)$ in this order then it is \emph{self-intersecting}.

\section{A Partitioned $2$-page Book-Embedding Testing Algorithm}\label{se:book-embedding}

In this section we describe an algorithm that, given
an instance of \be, decides whether it is positive and, in case it is,
constructs a book embedding of the input graph such that each edge is drawn
on the page it is assigned to. The algorithm is the one proposed
in~\cite{hn-tpbesg-09tr}. However,
substantial modifications have been applied to implement it. Part of them
aim at simplifying the algorithm, while others at decreasing the value of
some
constant factors spoiling the efficiency. Further, some
steps that are described at high level in~\cite{hn-tpbesg-09tr} are here
detailed. The main differences with~\cite{hn-tpbesg-09tr} are
highlighted throughout the paper.

Let $G(V,E_1,E_2)$ be an instance of problem \be. We say that
the edges of $E_1$ (of $E_2$) are \emph{red} (\emph{blue}) edges.
As pointed out in~\cite{hn-tpbesg-09tr}, 
the cases in which $G$ is disconncted or simply connnected can be easily 
 reduced to the case in which $G$ is biconnected, in the sense that 
$G$ admits a \be if and only if all
the biconnected components of $G$ admit a solution. 
 In fact, simply connected components can just be placed one after the other on the
 spine of the book embedding, while biconnected components need to be
 connected through their cut-vertices. However, it is easy to see that if a
 biconnected component admits a book embedding, then it admits a book
 embedding in which the cut-vertex connecting it to its parent component in
 the BC-tree is incident to the outer face. Namely, such a book embedding
 can be obtained by circularly rotating the vertices on the spine. Hence,
 it is always possible to merge the biconnected components  on the spine
 through their cut-vertices. 
Hence, we limit the description to the case in
which $G$ is biconnected. Moreover, we assume that both $E_1$ and $E_2$ are
not empty, since a graph with only red (blue) edges is a positive instance
if and only if it is outerplanar, which is testable in linear time.

\begin{figure}[bt]
  \centering{
\begin{tabular}{c c c c}
\mbox{\epsfig{figure=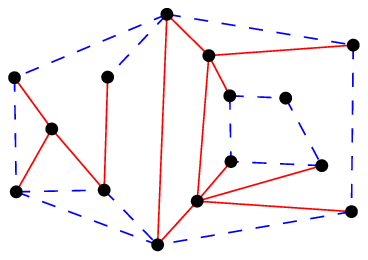,scale=0.6,clip=}}
\hspace{0.6cm} &
\mbox{\epsfig{figure=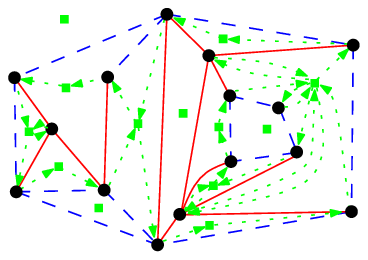,scale
=0.6,clip=}} 
\hspace{0.6cm} &
\mbox{\epsfig{figure=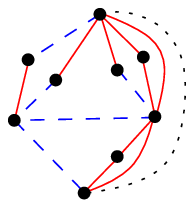,scale=0.8,clip=}}
\hspace{0.6cm} &
\mbox{\epsfig{figure=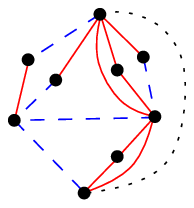,scale=0.8,clip=}}\\
(a) \hspace{0.6cm} & (b) \hspace{0.6cm} & (c) \hspace{0.6cm} & (d)
\end{tabular}}
\caption{(a) A disjunctive and splitter-free embedding of a graph. (b)
The corresponding green graph. (c) An r-rimmed embedding of a graph $G$.
(d) An embedding of $G$ that is not r-rimmed.}
  \label{fig:disjunctive-splitterfree}
\end{figure}

The algorithm is based on a characterization proved
in~\cite{hn-tpbesg-09tr} stating that an instance admits
a solution if and only if $G$ admits a \emph{disjunctive} and
\emph{splitter-free} planar embedding (see
Fig.~\ref{fig:disjunctive-splitterfree}(a)). An
embedding is \emph{disjunctive} if for each vertex $v \in V$ all the red
(blue) edges incident to $v$ appear consecutively around $v$. 
 Notice that, in the upward planarity literature, 
 disjunctive embeddings are often called
 \emph{bimodal}\cite{gt-upt-95}. 
A \emph{splitter} is a cycle $C$ composed of
red (blue) edges such that both the open regions of the
plane determined by $C$ contain either a vertex or a blue (red) edge. An
embedding is \emph{splitter-free} if it has no splitter. 
The first part of the algorithm, that is based on the SPQR-tree
decomposition of $G$ and whose details are in
Sections~\ref{se:spqr} and~\ref{se:nodes}, concerns the construction of an embedding of
$G$ satisfying these requirements, if it exists. Otherwise, $G$ does not
admit any solution.

Once a disjunctive and splitter-free embedding $\Gamma$ of $G$ has been
computed, an auxiliary graph $G^*$, called \emph{green graph}, is
constructed starting from $\Gamma$.
Then, as proved
in~\cite{hn-tpbesg-09tr}, a \be of $G$ can be constructed by computing a
non-self-intersecting Eulerian tour on $G^*$ and by placing the vertices of
$V$ on the spine in the order they appear on such a Eulerian tour.

Graph $G^*$ is a directed graph whose vertices are the vertices of $V$ plus
a vertex for each face of $\Gamma$. See
Fig.~\ref{fig:disjunctive-splitterfree}(b). 
 Edges of $G^*$ are determined
 as follows. 
For each vertex $v$ of $G$ incident to at least one
red edge and one blue edge, consider each face $f$ incident to $v$ such
that $v$ is between a red edge $e_1$ and a blue edge $e_2$ on $f$. If
$e_1$ immediately precedes $e_2$ in the clockwise ordering of the edges
around $v$, then add to $G^*$ an oriented edge $(v,f)$, otherwise add an
oriented edge $(f,v)$. For each
vertex $w$ of $G^*$ incident only to red (blue) edges, consider a face $f'$
incident to $w$ that contains at least one blue (red) edge. Since $\Gamma$
is splitter-free, such face exists. Then, add directed edges $(w,f')$ and
$(f',w)$.
Note that, by construction, $G^*$ is a bipartite plane digraph, every
vertex $v$ of $V$ has degree $2$ in $G^*$, namely $v$ is incident to
exactly one entering and one exiting edge, and each vertex $f$
corresponding to a face of $\Gamma$ has even degree, namely the number of
edges entering $f$ equals the number of edges exiting $f$, and such edges
alternate around $f$. From this and from the fact that the underlying
graph of $G^*$ is connected, as pointed out in~\cite{hn-tpbesg-09tr}, it
follows that $G^*$ contains a Eulerian tour. 

In the following we show that the alternation of entering and exiting edges around
each vertex ensures the existence of a non-selfintersecting Eulerian tour, as well.
In order to do that, we describe an algorithm that, given a disjunctive and splitter-free
embedding and the corresponding green graph $G^*$, computes a non-self-intersecting
Eulerian tour of $G^*$.

Given a plane embedded graph and an outer face $f$, we call \emph{boundary}
the set of (possibly non-simple) cycles composed of edges that are incident
to $f$. We proceed on the green graph $G^*$ as follows. Starting from any
outer face $f$ we iteratively remove at each step $i$ the edges of the
boundary $B_i$, thus identifying a new outer face and a new boundary, until
the graph is empty. On the cycles belonging to the extracted boundaries a
hierarchical relationship is defined as follows. Given two consecutive
boundaries $B_h$ and $B_{h+1}$, a cycle $C_j$ of $B_{h}$ is
the \emph{father} of a cycle $C_k$ of $B_{h+1}$ if $C_j$ and $C_k$ share a
vertex. This hierarchy can be easily represented by a tree, which we call
the \emph{boundaries tree}, whose nodes are the cycles of the boundaries
and whose root is the cycle representing the outer face of $G^*$. Given the
alternance of outgoing and incoming edges on the nodes of $G^*$, it is easy
to see that every connected component of a boundary $B_i$ is a directed
cycle. A Eulerian non-self-intersecting tour of $G^*$ is obtained by
visiting every cycle of the boundaries according to its orientation in the order
induced by a DFS visit of the boundaries tree. Namely, starting from an
edge of the cycle that is the root of the boundary tree, we construct
the tour by following the orientation of the edges. When a node $v$ of
degree greater than $2$ is encountered coming from an oriented edge
$(u,v)$ of a cycle, we start visiting its child cycle by following the
edge $(v,w)$ following $(u,v)$ in the clockwise order of the
edges around $v$. Note that, because of the alternance of entering and
exiting edges, edge $(v,w)$ is directed from $v$ to $w$. The same happens
when the visit of the child is finished and the visit of the father
continues. Hence, intersections in the Eulerian tour are always avoided.

From the above discussion it follows the claimed statement that the
described algorithm computes a \be of $(V,E_1,E_2)$, if any such a \be
exists. 

\section{Computing a Disjunctive and Splitter-Free Embedding}\label{se:spqr}

Let $G(V,E_1,E_2)$ be a biconnected planar graph. We describe an algorithm
to compute a disjunctive and splitter-free embedding of $G$, if any such an
embedding exists, consisting of two preprocessing
traversals of the SPQR-tree $\cal T$ of $G$ and of a final bottom-up
traversal to compute the required embedding.

Let $\mu$ be a node of $\cal T$. According to~\cite{hn-tpbesg-09tr}, a
virtual edge $e$ of \skel($\mu$) is an \emph{r-edge} (a
\emph{b-edge}) if there exists a path in \pert($\mu$) between the poles of
$\mu$ composed of red edges (of blue edges). If $e$ is both an r-edge and a
b-edge, it is a \emph{br-edge}. 

Consider a cycle $C = e_1, \dots, e_q$ in \skel($\mu$) composed of edges of
the same color, say r-edges. If $C$ is a splitter in every embedding of
\skel($\mu$), then a splitter is unavoidable. However, even
if there exists an embedding of \skel($\mu$) such that $C$ is not a
splitter, then a cycle in \pert($\mu$) passing through the pertinent graphs
of $e_1, \dots, e_q$ could still be a splitter (since $e_1, \dots, e_q$ are
r-edges, there exists at least one red cycle $C'$ in \pert($\mu$)). 
Consider any node $\nu$ corresponding to a virtual edge $e_i$ and the
path $p_\nu(C')$ between the poles of $\nu$ that is part of $C'$.
Intuitively, in order for $C'$ not to be a splitter, we should construct an
embedding of \pert($\nu$) in which $p_\nu(C')$ is on the outer face.
Actually, not all the vertices of $p_\nu(C')$ have to be on the outer face,
since red chords might exist in $p_\nu(C')$ (that is, red edges connecting
vertices not consecutive in $p_\nu(C')$), separating some vertex of
$p_\nu(C')$ from the outer face, as in this case such chords
would be internal to $C'$, and this does not make it a splitter.
 On the other hand, if $p_\nu(C')$ has a blue edge 
  or a vertex (even if this vertex belongs to another 
 path between the poles composed of red edges)
 on both its sides, then $C'$ becomes a splitter.
In analogy with~\cite{hn-tpbesg-09tr}, where the same concept was described
with a slightly different definition, we say that an embedding of
\pert$(\nu)$ in which each path between the poles composed of red edges (of
blue edges) has only red edges (blue edges) on one of its sides is
\emph{r-rimmed} (is \emph{b-rimmed}).
Figs.~\ref{fig:disjunctive-splitterfree}(c) and~(d) show an r-rimmed and a
non-r-rimmed embedding, respectively. Note that an embedding could be at
the same time both r- and b-rimmed, with the red and the blue paths on
different sides of the outer face. 

The existence of an r-rimmed (b-rimmed) embedding is necessary only
for each node $\mu$ such that there exists a cycle $C$ of red
(blue) edges traversing both $\mu$ and its parent. However,
the existence of $C$ is not known when processing $\mu$ during
a bottom-up visit of $\cal T$. Thus, we perform a preprocessing phase to
decide for each node $\mu$ whether any such cycle $C$ exists. In this
case, $\mu$ is \emph{r-joined} (\emph{b-joined}). Hence, when
processing $\mu$, we know whether it is r-joined (b-joined) and, in case,
we inductively compute an r-rimmed (b-rimmed) embedding.

\begin{figure}[bt]
  \centering{
\begin{tabular}{c c c c c c}
    \mbox{\epsfig{figure=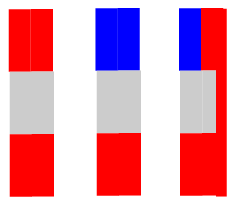,scale=0.5,clip=}}
\hspace{0.3cm} &
    \mbox{\epsfig{figure=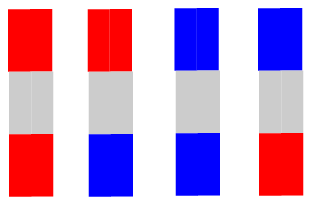,scale=0.5,clip=}}
\hspace{0.3cm} &
    \mbox{\epsfig{figure=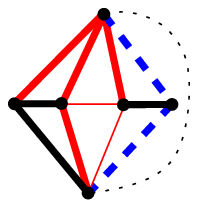,scale=0.7,clip=}} 
\hspace{0.3cm} &
    \mbox{\epsfig{figure=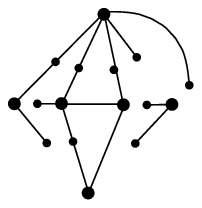,scale=0.7,clip=}} 
\hspace{0.3cm} &
    \mbox{\epsfig{figure=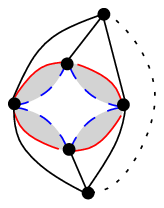,scale=0.7,clip=}}
\hspace{0.3cm} &
\mbox{\epsfig{figure=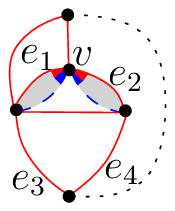,scale=0.7,clip=}}\\
(a) \hspace{0.3cm} & (b) \hspace{0.3cm} & (c)\hspace{0.3cm} &
(d) \hspace{0.3cm} & (e) \hspace{0.3cm} & (f)
\end{tabular}}
\caption{Parallel virtual edges are sketched with rectangles colored
according to their poles. (a) An r-rimmed embedding forces an $RBR$
color-pattern on a pole. (b) A color-pattern $BR$ or $RB$ on a pole forces
either an $RBR$ or a $BRB$ on the other pole. (c) An R-node. Virtual edges
representing Q-nodes are thin. (d) The corresponding auxiliary graph $O_1$.
(e) A splitter that is not a rigid-splitter. (f) Disjunctiveness
constraints on nodes $e_1$ and $e_2$ determine a splitter
$(e_1,e_2,e_3,e_4)$.}
  \label{fig:rimmed}
\end{figure}

Concerning disjunctiveness, for each vertex $w$ of \skel($\mu$) we have to
check whether the ordering of the edges around $w$ determined by the
embedded pertinent graphs of the child nodes incident to $w$ makes it
disjunctive.
In order to classify the possible orderings of edges around the poles of
a node we define, in analogy with~\cite{hn-tpbesg-09tr}, the
\emph{color-pattern} of a node $\mu$ on a vertex $v$ as the sequence of
colors of the edges of \pert($\mu$) incident to $v$. Namely, the color-pattern of
$\mu$ on $v$ is one of $R, B, RB, BR, RBR, BRB$.
Note that, if the color-pattern is either $R$ or $B$, then it is the same
in any embedding. Otherwise, it depends on the chosen embedding. Hence, it
might be influenced by the fact that the embedding needs to be r- or
b-rimmed (see Fig~\ref{fig:rimmed}(a)) and by the need of a particular
color-pattern on the other pole (see Fig~\ref{fig:rimmed}(b)).
Thus, a color-pattern either $RBR$ or $BRB$ could be forced on a pole $u$
of $\mu$ although an $RB$ or a $BR$ pattern would be possible as well.
Another factor influencing the color-pattern on $u$ is the presence of red
or blue edges incident to $u$ in the pertinent of
the parent $\nu$ of $\mu$. In fact, if $u$ has color-pattern $RBR$
($BRB$) and there is a blue (red) edge in \pert$(\nu)$ incident to $u$,
then $u$ is not disjunctive.
Thus, in the preprocessing phase we also determine
two flags for each pole $u$ of $\mu$, stating whether $\nu$ contains at
least one red (blue) edge incident to $u$. Hence, when processing $\mu$, we
know whether it is admissible to have an $RBR$ (a $BRB$) color-pattern on
its poles. 

 Hence, after the preprocessing phase, we can assume to know for
 each node $\mu$ the following information:

 \begin{enumerate}
  \item two flags stating whether $\mu$ is r-joined and whether it is
 b-joined;
  \item for each pole $u$ of $\mu$, two flags stating whether the parent
 $\nu$ of $\mu$ contains at least one red edge and whether it contains at
 least one blue edge, respectively, incident to $u$.
 \end{enumerate}

The two information obtained in the preprocessing can be properly combined
when processing a node to decide whether an embedding satisfying
all the constraints exists, as described is Section~\ref{se:nodes}.
If it is not the case, we state that the instance
is negative, while in the case that at least one of such embeddings exists,
we can arbitrarily choose one of them, without the need of carrying on a
multiplicity of embeddings.
This is one of the most crucial differences between our
implementation and ~\cite{hn-tpbesg-09tr}. In fact, even
if they perform a preprocessing to determine whether a node is
r-joined (b-joined), they do not exploit it for disjunctiveness, and have
to consider at each step all the possible embeddings
determining different color-patterns on the two poles. Of course, as the
number of color-patterns is bounded by a
constant, this does not affect the asymptotic complexity, but our
solution noticeably improves on the execution times. Also, they deal with
constraints given by the r-joinedness (b-joinedness) and by the
disjunctiveness in two different steps. 
In our case, instead, instances that are negative due to disjunctiveness
are recognized much earlier.

The preprocessing consists of a bottom-up and a top-down
traversal of $\cal T$, that we describe in the following. 
The bottom-up traversal computes some information on each node, which are then
aggregated in the top-down traversal to efficiently compute the needed information on the
parent of each node.

In order to determine which are the r- and the b-joined nodes, in
the bottom-up traversal we compute for each node whether its skeleton
(excluding the virtual edge representing the parent) contains a path
between its poles composed of r-edges (b-edges). Then, in the top-down
traversal we transmit this information from each node to its children,
namely all and only the children that are part of a cycle composed of
r-edges (b-edges) in the skeleton of a node are r-joined (b-joined).

In order to determine which are the nodes whose parent has at least a red
(a blue) edge incident to a pole $u$, we determine for each node $\mu$
in the bottom-up traversal whether it contains a red (blue) edge
incident to $u$, and in case it does, we add $1$ to a counter associated
with $u$ and the parent of $\mu$. 
Then, during the top-down traversal we inductively compute the information
on each node $\mu$, we accordingly update the counter associated with $u$
and $\mu$ for each child node of $\mu$, and we state that the parent of a
child node $\nu$ of $\mu$ has a red (blue) edge incident to $u$ either if
the value of the counter is at least $2$ or if it is $1$ and $\nu$ has no
red (blue) edge incident to $u$.

\remove{
\textcolor{red}{[anche rileggendola a mente fredda continua a
sembrarmi poco
chiara. Provo a dare la mia versione. Then, during the top-down traversal
we inductively compute the information on each node $\mu$. The counter
associated to a pole $u$ and the parent of $\mu$ is updated accordingly to
the value of the parent of $\mu$: we add $1$ to the red (blue) counter if the
parent has a red (blue) edge incident to $u$. Then, for each
child node $\nu$ of $\mu$, we state that the parent of $\nu$ has a red (blue)
edge incident to $u$ either if
the value of the counter is at least $2$ or if it is $1$ and $\nu$ has no
red (blue) edge incident to $u$.}
}

 In the next section we describe the final bottom-up 
 traversal of $\cal T$ which computes a disjunctive and splitter-free embedding of $G$, if
it exists.

\section{SPQR-tree Algorithm}\label{se:nodes}

When considering a node $\mu$ of $\cal T$ with children $\nu_1,
\dots, \nu_k$, exploiting the information resulting from the
preprocessing and the information inductively computed for
$\nu_1, \dots, \nu_k$, we check whether $\mu$ admits a splitter-free and
disjunctive embedding and compute the following: (i) if
$\mu$ is r-joined (b-joined), an r-rimmed (a b-rimmed) embedding; and (ii)
the color-patterns of the poles of $\mu$.

\remove{
\textcolor{red}{Mi sembra impreciso. Mia versione:
When considering a node $\mu$ of $\cal T$ with child components $\nu_1,
\dots, \nu_k$, we check whether $\mu$ admits a splitter-free and
disjunctive embedding. Namely we look for an embedding:
\begin{enumerate}
 \item that is r-rimmed (b-rimmed) if $\mu$ is r-joined (b-joined);
 \item in which the sequence of color-patterns of the virtual edges incident to a node,
considered in clockwise order, represents a disjunctive sequence of incident edges.
\end{enumerate} 
In order to find such an embedding we use the following information inductively computed for $\nu_1, \dots,
\nu_k$, together with the information resulting from the preprocessing phase:
\begin{enumerate}
 \item whether $\nu_i$ is r-rimmed (b-rimmed);
 \item the color-patterns associated to the poles of $\nu_i$.
\end{enumerate}
}
}

In the base case, $\mu$ is a \textbf{Q-node}. Suppose that \skel($\mu$) is
an r-edge, the other case being analogous. If $\mu$ is r-joined, every
embedding of \skel($\mu$) is r-rimmed. Further,
the color-pattern on the poles is $R$ in any embedding of \skel($\mu$).

Suppose that $\mu$ is an \textbf{R-node}. Since \skel$(\mu)$ is
triconnected, it has one planar embedding, up to a flip. Hence, if there is
a splitter in \skel($\mu$), then it is unavoidable. Hong and Nagamochi call
such splitters \emph{rigid-splitters}. 
In order to test the existence of such splitters, for each set $E_i$,
$i=1,2$, we construct an auxiliary graph $O_i$ starting from
\skel($\mu$). See Figs.~\ref{fig:rimmed}(c) and~(d). We
describe the construction for $E_1$, the other case being analogous.
Initialize $O_1 = $\skel($\mu$). Subdivide each virtual edge of
\skel($\mu$) (including the one representing the parent) with a dummy
vertex, except for the r-edges corresponding to Q-nodes. Then, for each
dummy vertex subdividing a virtual edge
that is not an r-edge, remove one of its incident edges without modifying
the embedding. Finally, check whether the obtained embedding of $O_1$ is an
outerplane embedding, that is, all the vertices of $O_1$ are on the same
face. 
This check is performed by iterating on all the faces of the
embedded graph $O_1$ and by checking whether there exists one containing
all the vertices. Note that this step can be performed in linear time,
since each vertex of degree $d$ is examined at most $d$ times and since the
sum of the degrees of the vertices of a graph is twice the number of edges,
which is $O(n)$.
In~\cite{hn-tpbesg-09tr} this step is performed by
constructing a variant of the green graph and checking whether it is
connected. Even if the time complexity of the two approaches is basically the same, we
find that our approach is easier to implement and slightly more efficient, since $O_1$
does not need to be constructed, but can be obtained by flagging the edges of
\skel$(\mu)$.

Note that, for each cycle composed of r-edges (b-edges) in \skel($\mu$)
that is not a rigid-splitter, all the nodes composing it inductively admit
an r-rimmed (b-rimmed) embedding. Hence, it suffices to flip them in such a
way that their red (blue) border is turned towards the red (the blue)
outerplanar face. However, if each of them has an embedding that is both
r-rimmed and b-rimmed, the red and the blue outerplanar faces coincide and
it is not possible to flip the nodes properly, which implies that a
splitter exists in the embedding. See Fig.~\ref{fig:rimmed}(e). This type
of splitter seems to have gone unnoticed in~\cite{hn-tpbesg-09tr}, where
flips imposed by cycles of r- and b-edges are considered independently. 

We deal with disjunctiveness constraints. We observe some
straightforward properties of the color-patterns of the nodes
incident to the same vertex $w$ of \skel($\mu$). (i) At most two nodes
have color-pattern different from $R$ and $B$. (ii) If one node
has color-pattern $RBR$ ($BRB$), then all the other
nodes have color-pattern $R$ ($B$). Hence, since each vertex
has degree at least $3$ in \skel($\mu$), at least one node $\nu$ incident
to $w$ exists with color-pattern either $R$ or $B$. Thus, starting
from $\nu$, we consider all the nodes incident to $w$ in clockwise order
and greedily decide a flip based on the current color. If more than two
changes of color are performed, then $G$ does not admit any
disjunctive embedding. If exactly one node
$\nu$ has color-pattern different from $R$ or $B$ and all the other nodes
have color-pattern $R$ ($B$), then the flip of $\nu$ is not decided at
this step. Also, the flip is not decided for the nodes having
color-pattern $R$ or $B$.

Disjunctiveness and splitter-free constraints might be in
contrast. See Fig.~\ref{fig:rimmed}(f). We can efficiently determine such
contrasts by flagging the nodes that need to be flipped and, in case such
contrasts exist, state that the instance is negative. This check is not
described in~\cite{hn-tpbesg-09tr}, where possible contrasts between
disjunctive and splitter-free constraints are noticed for P-nodes but not
for R-nodes.

The color-patterns of the poles and, if needed, an r-rimmed (a b-rimmed)
embedding of \pert($\mu$) are computed by considering the information
on the parent node, the color-patterns of the virtual edges incident to
the poles, and the r-rimmed (b-rimmed) embedding of the children.

Suppose that $\mu$ is an \textbf{S-node}. Since \skel($\mu$) is a cycle
containing all the virtual edges, even if such a cycle is composed of edges
of the same color, then it is not a splitter. Namely, even if there exist
both a red and a blue cycle passing through all the children of $\mu$, such
nodes can be flipped so that the red and the blue borders are turned
towards the two faces of \skel($\mu$).

Concerning disjunctiveness constraints, if two children both
incide on a vertex $u$ of \skel$(\mu)$ with color-pattern either
$BR$ or $RB$, then they have to be flipped in such a way that the red edges
(and hence the blue edges) are consecutive around $u$. In all the other
cases, the relative flip of the two children incident to $u$ is not fixed
by their color-patterns. If there exists at least a vertex $u$ with this
property, we say that $\mu$ admits two different \emph{semi-flips}.
Intuitively, this means that the color-pattern of a pole is
independent of the one on the other pole, since they depend on flips
performed on two different subsets of children of $\mu$.

Note that in an S-node no contrast between splitter-free and
disjunctiveness
constraints are possible, since flipping the r-rimmed embeddings
towards the same face implies placing the red edges consecutive around $u$.
Hence, no negative answer can be given during the processing of an S-node.

The color-pattern on each pole is the color-pattern of the unique
node incident to it, while an r-rimmed (b-rimmed) embedding is obtained by
concatenating the r-rimmed (b-rimmed) embeddings of the children.

Suppose that $\mu$ is a \textbf{P-node}. In order for a splitter-free
embedding
to exist, the following must hold:
\begin{inparaenum}[(i)]
 \item There exist at most $3$ r-edges (b-edges); if they are $3$ then
one is a Q-node.
 \item There exist at most $2$ virtual edges that are both r-edges and
b-edges; if they are $2$ then there exists only another virtual edge and
it is a Q-node.
\end{inparaenum}
When such conditions do not hold, the r-edges (b-edges) induce a
splitter in every embedding of the P-node. 
\remove{
Otherwise they can be grouped (and treated as a
single component) so that the r-rimmed (b-rimmed) edges are adjacent in the edge
permutation of the P-node. This reduces to one the number of virtual edges that
have a colored path between the poles.
}

On the other hand, in order for a disjunctive embedding to exist,
the following must hold:
\begin{inparaenum}[(i)]
 \item if there exists a virtual edge with $RBR$ ($BRB$) color-pattern on a pole,
then all the other edges have color-pattern $R$ ($B$) on that pole;
 \item there exist at most two virtual edges with color-pattern $RB$ or $BR$ on a pole.
\end{inparaenum}
When these conditions do not hold for a pole $u$, in every embedding of $G$ there exist
more than two color changes in the clockwise ordering of edges incident to $u$, that
is, there exists no embedding that makes $u$ disjunctive.

Consider a child node $\nu_1$ having color-pattern either $R$ or $B$ on
both poles, say $R$ on pole $u$ and $B$ on pole $v$, and consider another
child node $\nu_2$ having color-pattern $R$ on $u$ and $B$ on $v$. Nodes
$\nu_1$ and $\nu_2$ can be considered as a single node $\nu^*$ with
color-patterns $R$ and $B$ on the two poles. When the permutation of the
P-node has been computed, $\nu^*$ is replaced by $\nu_1$ and $\nu_2$.
This operation reduces the number of virtual edges to at most
$8$, namely at most $4$ groups of nodes having either $R$ or $B$ on
both poles plus at most $2$ nodes with color-pattern different from $R$ and
$B$ on a pole and at most $2$ nodes with color-pattern different from $R$
and $B$ on the other pole. Note that the parent cannot be grouped, since
its color-patterns are unknown at this stage.
In \cite{hn-tpbesg-09tr} this fact is exploited to search an embedding
with the desired properties by exhaustively checking all permutations,
i.e., with a brute-force approach. However,
even if the time complexity is aympthotically linear, this yields a
huge number of cases, namely $8! * 2^8$ combinations, i.e., all
permutations of $8$ edges multiplied by all flip choices.

\begin{figure}[tb]
  \centering{
\begin{tabular}{c c}
    \mbox{\epsfig{figure=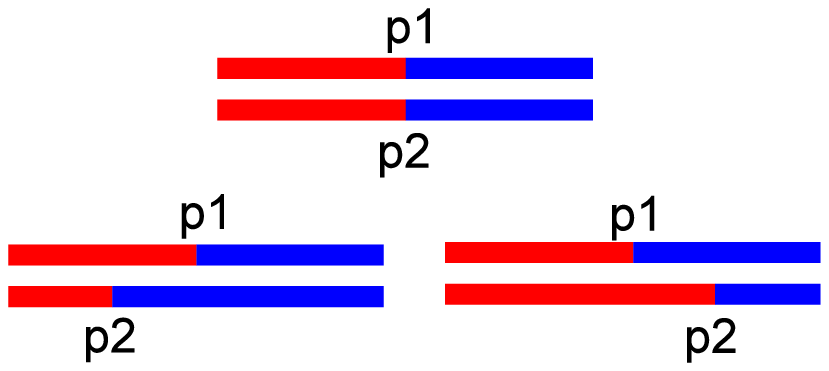,scale=0.37,clip=}}
\hspace{1cm} &
    \mbox{\epsfig{figure=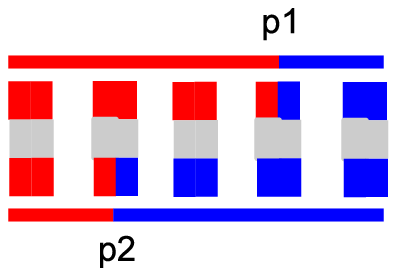,scale=0.55,clip=}}\\
    (a) \hspace{0.5cm} & (b)
\end{tabular}}
\caption{
(a) All possible alignments of a pair of $RB$ color-patterns. (b)
Correspondence between an alignment of a pair of color-patterns and a
sequence of virtual edges of a P-node.}
\label{fig:alignments}
\end{figure}

Hence, our implementation uses a different approach in order to search into
a much smaller space.
Namely, consider any color-pattern, say $RBR$, and map it to a
linear segment of fixed length, partitioned into
three parts $R$, $B$, $R$, by two points that represent the two changes of
color $R-B$ and $B-R$. Such points are identified by a unidimensional
coordinate $p$ along the segment. Given two color-patterns,
their segments, and a separating point for each of them, with
coordinates $p_1$ and $p_2$, respectively, any of the following conditions
can hold:
\begin{inparaenum}[(i)]
 \item $p_1 < p_2$;
 \item $p_1 = p_2$;
 \item $p_1 > p_2$.
\end{inparaenum}
See Figure~\ref{fig:alignments}(a). We call \emph{alignment} of two
color-patterns each combinatorial possibility obtained by exhaustively
making conditions (i)-(iii) hold for all pairs of separating points
of their segments.
An alignment of two color-patterns $P_1,P_2$ uniquely corresponds to a sequence of virtual edges
whose color-patterns compose $P_1$ and $P_2$ on the two poles. Such sequence makes both poles
disjunctive by construction. See Fig.~\ref{fig:alignments}(b).
Our approach exploits this fact by exhaustively
enumerating all alignments of all pairs of color-patterns. The result is the set $L$
containing all and only the disjunctive edge permutations of a generic P-node. $L$ contains
exactly $180$ elements.
The pseudocode of the algorithm
{\sc Generate\_Admissible\_Solutions\_for\_P-nodes} performing the enumeration that
constructs $L$ is given in Algorithm~\ref{alg:genera-soluzioni-ammissibili-per-nodo-p}.

\begin{algorithm}[!htb]
\caption{\sc{Generate\_Admissible\_Solutions\_for\_P-nodes}}
\begin{algorithmic}[1]
\algsetup{indent=1em,linenosize=\small,linenodelimiter=.}
\label{alg:genera-soluzioni-ammissibili-per-nodo-p}
\medskip

\STATE $L \gets$ empty list of permutations of virtual edges
\STATE $S \gets$ list of color-patterns: $R$, $B$, $RB$, $BR$, $RBR$, $BRB$
\FORALL {$\sigma_1 \in S$}
	\FORALL {$\sigma_2 \in S$}
		\FORALL {alignment of $\sigma_1$ and $\sigma_2$}
			\STATE $Z \gets$ list of pairs of colors, where
each pair is composed of a color of $\sigma_1$ and of a color of
$\sigma_2$. Elements of the list are determined by discretizing the
alignment. Note that each alignment determines at most nine list elements.
			\STATE $P \gets$ empty list of virtual edges
			\FORALL {$z \in Z$}
				\STATE append to $P$ a new edge $p$
with color-patterns on its poles $\in \{R, B\}$ corresponding to the
colors of $z$
				\IF {the color-patterns of the poles of 
$p$ are either both $R$ or both $B$}
					\STATE make $p$ r-rimmed or
b-rimmed depending on whether the color-patterns are $R$ or $B$
				\ENDIF
				\IF {$p$ is the first or the last element
of $Z$}
					\STATE flip $p$ in such a
way that the r-rimmed (b-rimmed) path is towards outside
				\ENDIF
			\ENDFOR
			\FORALL {$p \in P$}
				\IF {the first color of the color-pattern
of $p$ on a pole is different from the last color of the color-pattern of
the edge preceding $p$ in $P$ on the same pole}
					\STATE insert a new edge $p'$
preceding $p$ with color-pattern $RB$ or $BR$ on the considered pole 
				\ENDIF
				\IF {the color-patterns of $p'$ on the
two poles have the same first (last) color}
					\STATE make $p'$ r-rimmed or
b-rimmed
				\ENDIF
			\ENDFOR
			\STATE $D \gets$ edges that have color-pattern $R$
or $B$ on both poles and multiple instances in $P$
			\IF {size($D$) = $0$}
				\STATE append($P$, $L$)
			\ELSE
				\FORALL {$p \in D$}
					\FORALL {instance $p_i$ of $p$ in $P$}
						\STATE $P' \gets$ copy of
P with only instance $p_i$ of $p$
						\STATE append($P'$, $L$)
					\ENDFOR
				\ENDFOR
			\ENDIF			
		\ENDFOR	
	\ENDFOR
\ENDFOR
\RETURN $L$

\end{algorithmic}
\end{algorithm}

Since the virtual edges of a P-node have a disjunctive permutation if and
only if they can be disposed in the same sequence as an element in $L$,
a disjunctive embedding can be found, if it exists, by a brute
force search across the $180$ elements of $L$, an impressive
improvement with respect to the algorithm in \cite{hn-tpbesg-09tr}.

As the parent node could not be grouped with other nodes, it could impose
some additional constraints on the permutation to find that forbid
permutations having color-patterns $RBR$ or $BRB$ and that require any
r-rimmed (b-rimmed) node to be either the first or the last, if
the P-node is r-joined (b-joined).

The whole P-node algorithm must be repeated for every possible choice of semi-flip
for the virtual edges admitting it. However, at most two such virtual
edges can exist, since they have color-patterns $RB$ or $BR$ on
both poles. Hence, the algorithm must be repeated up to $4$ times.

\section{Experimental Results}\label{se:experiments}

In this section we describe the experimental tests performed to check 
correctness and efficiency of our implementation.
When performing experiments a crucial aspect is to have at disposal a
representative set of negative and positive instances. Negative
instances have the main role of checking the correctness, while
positive instances are both used to check the correctness and to test the
performance in a complete execution, without being influenced by early
recognition of negative instances.
We constructed the former set using ad-hoc examples, conceived to stress
all the steps of the algorithm. 
On the other hand, in order to obtain a suitable set of positive instances, we used random
generation. Unfortunately, to the best of our knowledge, no graph generator is
available to uniformly create graphs with a \be. Hence, we devised and
implemented a graph generator, whose inputs are a number $n$ of
vertices and a number $m \leq 3n - 6$ of edges. The output is a positive
instance of \be selected uniformly at random among the positive instances
with $n$ vertices and $m$ edges.

The generator works as follows. First, we place $n$ vertices
$v_1,\dots,v_n$ on the spine in this order. Then we
insert, above (below) the spine, red (blue) dummy edges $(v_1,v_2), \dots,
(v_{n-1},v_n)$, and $(v_1,v_n)$. In this way we initialize the two pages
with two faces $(v_1,\dots,v_n)$ composed of red and of blue dummy edges,
respectively. Observe that we inserted multiple dummy edges. Dummy edges
will be removed at the end. 
Second, we randomly select a face $f$ with at least three vertices,
selected with a probability proportional to the number of candidate edges
that can be added to it.
Then, an edge $(u,v)$ is chosen uniformly at random
among the potential candidate edges of $f$.
Edge $(u,v)$ is added to $f$ by either splitting $f$ or substituting a
dummy edge of $f$ with a ``real'' edge. Edge $(u,v)$ is colored red or blue
according to the color of the edges of $f$.
If $(u,v)$ is red (blue) we check if there exists a blue (red) face that
contains both $u$ and $v$ and remove $(u,v)$ from the candidate edges of
that face.
We iteratively perform the second step until $m$ is reached and at the end
we remove the dummy edges that have not been substituted by a ``real''
edge. Observe that in this way we do not generate multiple edges and that
the generated graphs are not necessarily connected.

We generated three test suites, Suite $1$, $2$, and $3$, with $m=2
n$, $m=2.5 n$, and $m=3n-6$, respectively.
For each Suite, we constructed ten buckets of instances, ranging from
$n=10,000$ to $n=100,000$ with an increment of $10,000$ from one
bucket to the other. For each bucket we constructed five instances with the
same parameters $n$ and
$m$. 
The choice of diversifying the edge density is motivated by the wish of testing the
performance of the algorithm on a wide variety of SPQR-trees, with Suite $3$ being a limit
case.

The algorithm was implemented in C++ with GDToolkit~\cite{dd-hgdvg-11}. The
OGDF library~\cite{OGDF} was used to construct the SPQR-trees. We used GDToolkit because
of its versatile and easy-to-use data structures and OGDF to construct SPQR-trees in
linear time.

Among the technical issues, the
P-node case required the analysis of a set of cases that is so large to
create correctness problems to any, even skilled, programmer. Hence, we
devised a code generator that, starting from a formal specification of the
constraints, wrote automatically the required C++ code.
For performing our experiments, we used an environment with the following
features:
\begin{inparaenum}[(i)]
 \item CPU Intel Dual Xeon X5355 Quad Core (since the algorithm is
sequential we used just one Core) 2.66GHz 2x4MB 1333MHz FSB.
 \item RAM 16GB 667MHz.
 \item Gentoo GNU/Linux (2.6.23).
 \item g++ 4.4.5.
\end{inparaenum}

\begin{figure}[bt]
  \centering{
\begin{tabular}{c c c}
    \mbox{\epsfig{figure=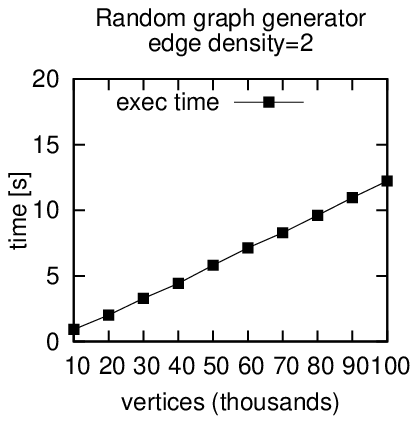,scale=1.0}} \hspace{-0.5cm}
& 
    \mbox{\epsfig{figure=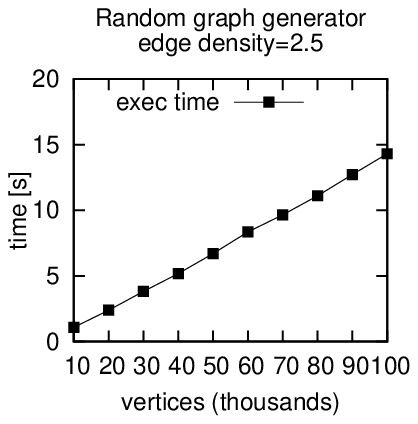,scale=1.0}} \hspace{-0.5cm}
&
    \mbox{\epsfig{figure=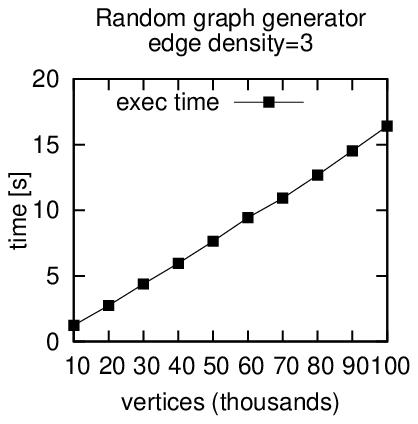,scale=1.0}}\\
    (a) \hspace{0.5cm} & (b) \hspace{0.5cm} & (c)
\end{tabular}}
\caption{Execution times of the generator for generating the three suites. The $x$-axis
represents the number of vertices of the bucket, while the $y$-axis represents the
average execution time on the instances in the bucket.}
  \label{fig:chart_generator}
\end{figure}

Figs.~\ref{fig:chart_generator}(a)--(c) show the execution times of the
generator for generating the three suites. 

Before giving the execution times of the algorithm on the generated instances, we show
some charts describing the structure of such instances, both in terms of connectivity and
in terms of the complexity of the corresponding SPQR-trees.

\begin{figure}[bt]
  \centering{
\begin{tabular}{c c c}
    \mbox{\epsfig{figure=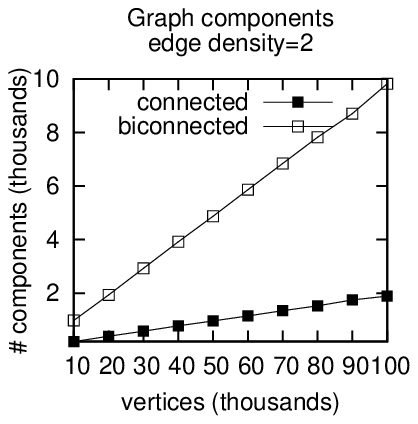,scale=1.0}} \hspace{-0.5cm}
& 
    \mbox{\epsfig{figure=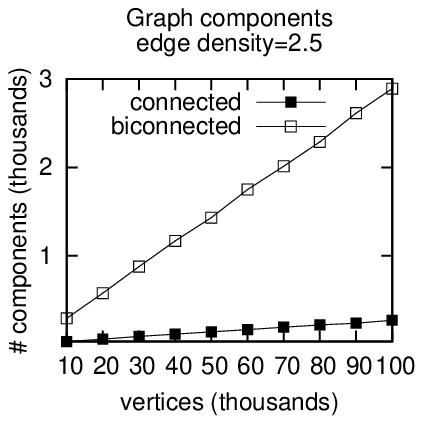,scale=1.0}} \hspace{-0.5cm}
&
    \mbox{\epsfig{figure=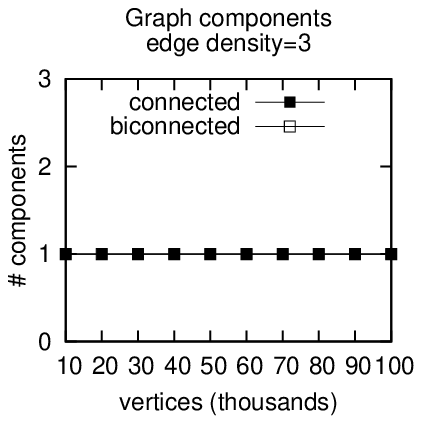,scale=1.0}}\\
    (a) \hspace{0.5cm} & (b) \hspace{0.5cm} & (c)
\end{tabular}}
\caption{The $x$-axis represents the number of vertices of the bucket, while the
$y$-axis represents the average number of components of each bucket.}
  \label{fig:chart_num_components}
\end{figure}

Figs.~\ref{fig:chart_num_components}(a)--(c) show the number of connected
(including isolated vertices) and biconnected (including single edges) components in the
test suites.

\begin{figure}[bt]
  \centering{
\begin{tabular}{c c c}
    \mbox{\epsfig{figure=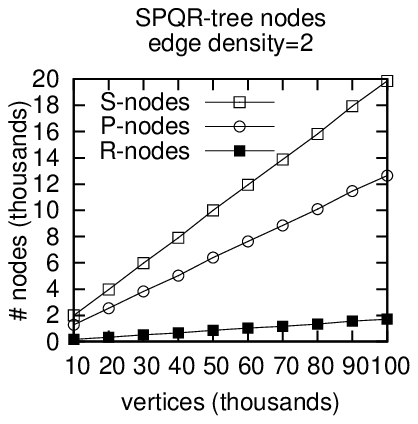,scale=1.0}} \hspace{-0.5cm}
& 
    \mbox{\epsfig{figure=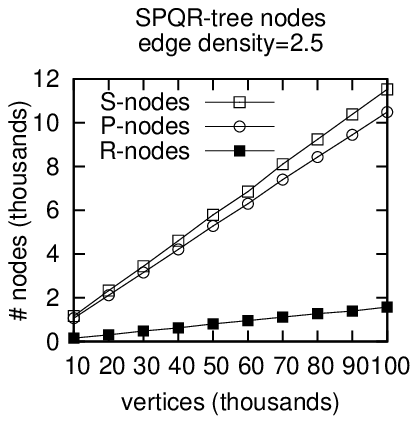,scale=1.0}} \hspace{-0.5cm}
&
    \mbox{\epsfig{figure=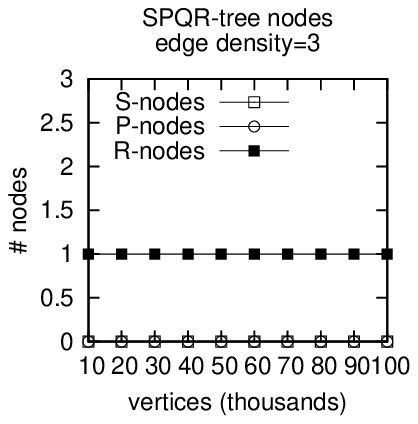,scale=1.0}}\\
    (a) \hspace{0.5cm} & (b) \hspace{0.5cm} & (c)
\end{tabular}}
\caption{Number of SPQR-tree nodes in the three test suites. The $y$-axis
represents the average number of SPQR-tree nodes of each bucket.}
  \label{fig:chart_num_spqr_nodes}
\end{figure}

Figs.~\ref{fig:chart_num_spqr_nodes}(a)--(c) show the number of 
SPQR-tree nodes in the three test suites. Note that the large amount of
P-nodes in Suites $1$ and $2$ puts in evidence how crucial has been in the
implementation to optimize the P-nodes processing.

Then, we give the execution times of the algorithm on such instances and an analysis of
them from several points of view.

\begin{figure}[bt]
  \centering{
\begin{tabular}{c c c}
    \mbox{\epsfig{figure=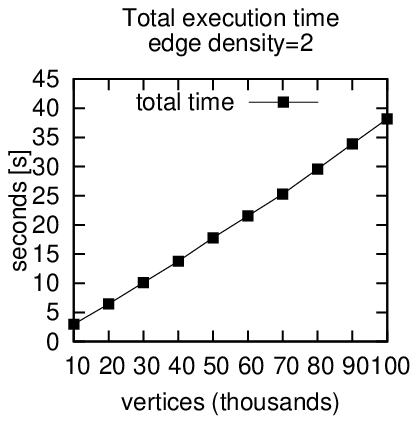,scale=0.95}}
\hspace{-0.5cm} & 
    \mbox{\epsfig{figure=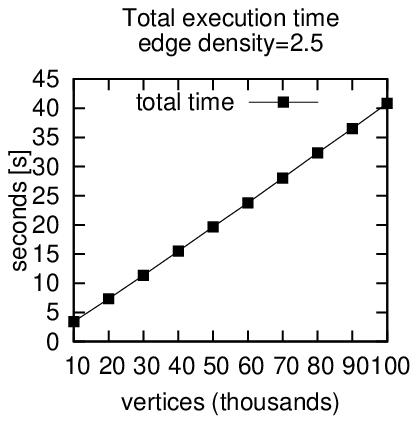,scale=0.95}}
\hspace{-0.5cm} &
    \mbox{\epsfig{figure=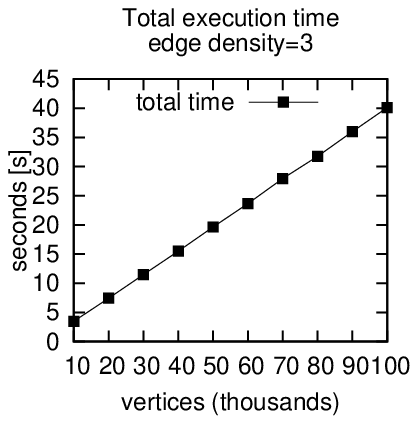,scale=0.95}}\\
    (a) \hspace{0.5cm} & (b) \hspace{0.5cm} & (c)
\end{tabular}}
\caption{Total execution time of the algorithm on the three test suites. The $y$-axis
represents the average execution time of the algorithm on the instances in the
bucket.}
  \label{fig:chart_total_execution_time}
\end{figure}

Fig.~\ref{fig:chart_total_execution_time}(a)--(c) show the total execution
times for the three suites.
These measurements include the time necessary to decompose the
graphs in their connected, biconnected, and triconnected components. The
algorithm clearly shows linear running times, with very little differences
among the three suites.

\begin{figure}[bt]
  \centering{
\begin{tabular}{c c c}
    \mbox{\epsfig{figure=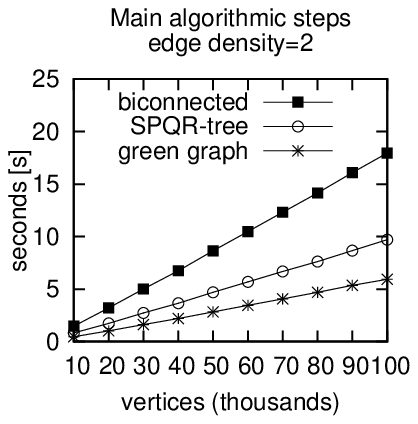,scale=1.0}} \hspace{-0.5cm} & 
    \mbox{\epsfig{figure=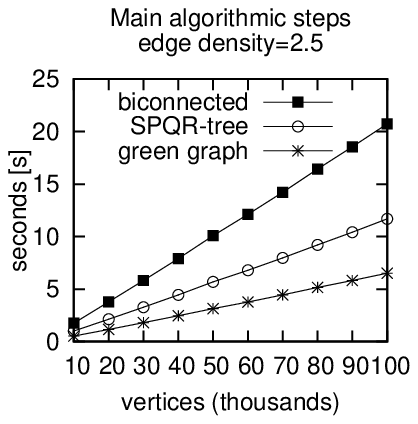,scale=1.0}} \hspace{-0.5cm} &
    \mbox{\epsfig{figure=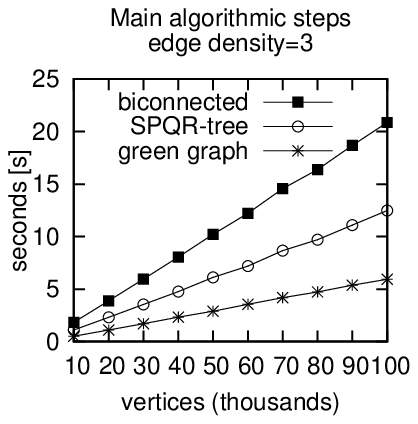,scale=1.0}}\\
    (a) \hspace{0.5cm} & (b) \hspace{0.5cm} & (c)
\end{tabular}}
\caption{Execution times of the algorithm for the biconnected components of the three test suites.}
  \label{fig:chart_main_alg_steps}
\end{figure}

Figs.~\ref{fig:chart_main_alg_steps}(a)--(c) show the execution times of
the main algorithmic steps for the three suites, namely (i) the total time spent to
process biconnected components, (ii) the time spent to deal with the SPQR-trees (excluding
the time to create them), and (iii) the time spent to
create the green graphs and to find the Eulerian tours.
Beside remarking the linear running time, these charts show how the time spent on
biconnected components is distributed among the two main algorithmic steps.

\begin{figure}[bt]
  \centering{
\begin{tabular}{c c c}
    \mbox{\epsfig{figure=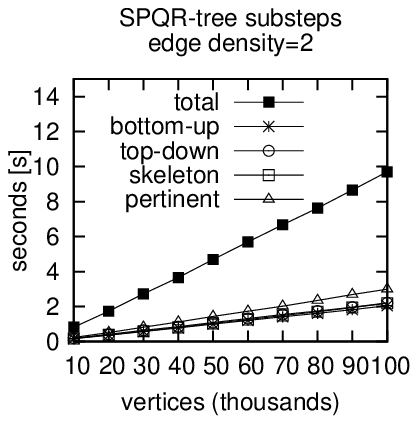,scale=1.0}} \hspace{-0.5cm}
& 
    \mbox{\epsfig{figure=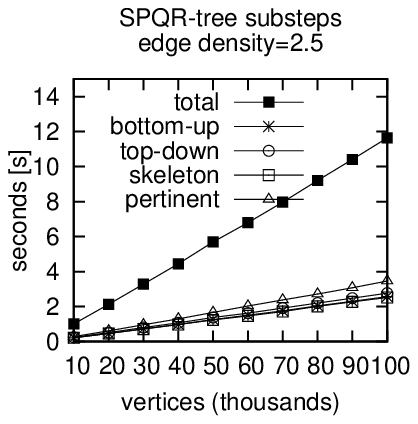,scale=1.0}} \hspace{-0.5cm}
&
    \mbox{\epsfig{figure=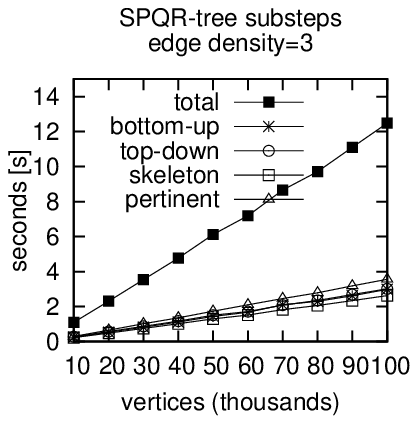,scale=1.0}}\\
    (a) \hspace{0.5cm} & (b) \hspace{0.5cm} & (c)
\end{tabular}}
\caption{The execution times of the four algorithmic substeps of the step
that deals with the SPQR-trees (excluding creation) for the three suites.}
  \label{fig:chart_spqr_steps}
\end{figure}

Figs.~\ref{fig:chart_spqr_steps}(a)--(c) show the execution times of the
four algorithmic substeps of the step that deals with the SPQR-trees
(excluding creation) for the three suites.
Namely, the five curves show: (i) the time to deal with the SPQR-trees
(excluding the time spent to create them),
(ii) preprocessing bottom-up phase,
(iii) preprocessing top-down phase,
(iv) the bottom-up skeleton embedding phase, and
(v) the pertinent graph embedding phase.

\begin{figure}[bt]
  \centering{
\begin{tabular}{c c c}
    \mbox{\epsfig{figure=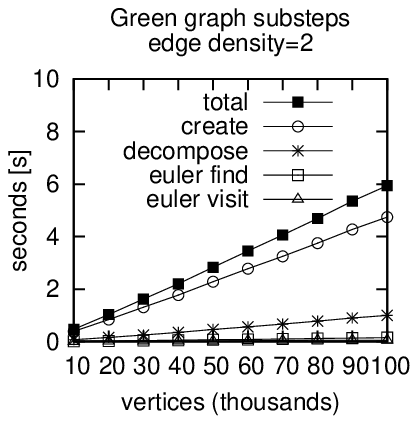,scale=1.0}} \hspace{-0.5cm}
& 
    \mbox{\epsfig{figure=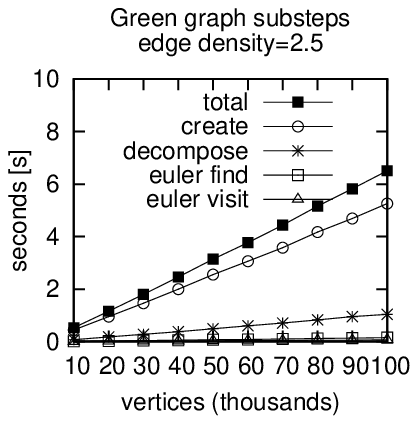,scale=1.0}}
\hspace{-0.5cm} &
    \mbox{\epsfig{figure=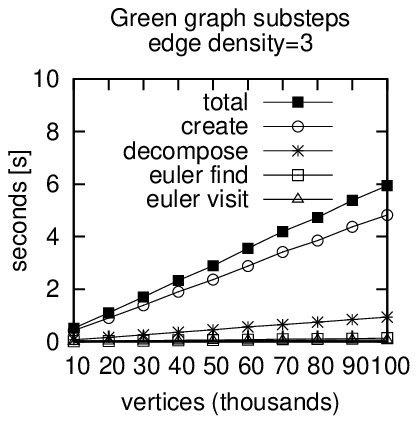,scale=1.0}}\\
    (a) \hspace{0.5cm} & (b) \hspace{0.5cm} & (c)
\end{tabular}}
\caption{The execution times of the
four algorithmic substeps of the step that
deals with the green graph for the three suites.}
  \label{fig:chart_green_steps}
\end{figure}

Figs.~\ref{fig:chart_green_steps}(a)--(c) show the execution times of the
four algorithmic substeps of the step that
deals with the green graph for the three suites.
Namely, the five curves show: (i) the time to deal with the green graph,
(ii) creation phase,
(iii) decomposition phase,
(iv) Eulerian tour finding phase, and
(v) Eulerian tour visiting phase.

\begin{figure}[bt]
  \centering{
\begin{tabular}{c c c}
    \mbox{\epsfig{figure=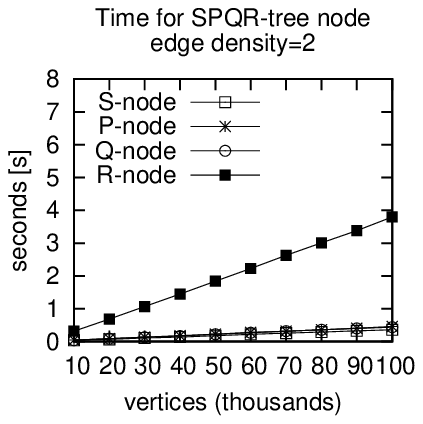,scale=1.0}}
\hspace{-0.5cm} & 
    \mbox{\epsfig{figure=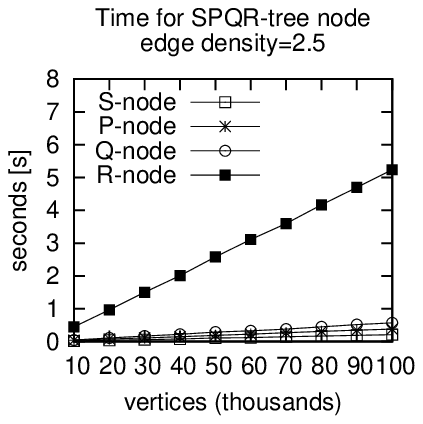,scale=1.0}}
\hspace{-0.5cm} &
    \mbox{\epsfig{figure=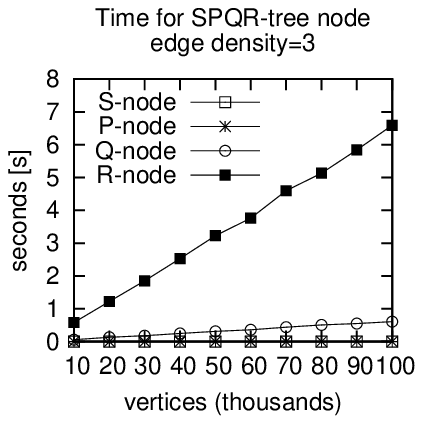,scale=1.0}}\\
    (a) \hspace{0.5cm} & (b) \hspace{0.5cm} & (c)
\end{tabular}}
\caption{The time spent to deal with the different types of SPQR-tree nodes
(excluding creation) for the three suites.}
  \label{fig:chart_spqr_nodes_time}
\end{figure}

Figs.~\ref{fig:chart_spqr_nodes_time}(a)--(c) show the time spent to deal
with the different types of SPQR-tree nodes (excluding
creation) for the three suites. Namely, the four curves show: (i) the
time for S-nodes, (ii) the time for P-nodes,
(iii) the time for Q-nodes, and (iv) the time for R-nodes.

\section{Conclusions}\label{se:conclusions}

We described an implementation of a constrained version of
the $2$-page book embedding problem in which the edges are assigned
to the two pages and the goal is to find an ordering of the vertices on the
spine that generates no crossing on each page.
The implemented linear time algorithm is the one given in~\cite{hn-tpbesg-09tr},
with several variations aimed at
simplifying it and at improving its performance.

We performed a large set of experimental tests on randomly generated
instances. From these experiments we conclude that the
original algorithm, together with our variations, correctly
solves the given problem, and that its performance are pretty good
on graphs of medium-large size.

\bibliographystyle{plain}
\bibliography{bibliography}

\begin{thebibliography}{10}

\bibitem{b-dppIIItcep-98}
T.~Biedl.
\newblock Drawing planar partitions {III}: {Two Constrained Embedding
  Problems}.
\newblock Tech. {Report} RRR 13-98, Rutcor Research Report, 1998.

\bibitem{Biedl:1998:DPP}
T.~C. Biedl, M.~Kaufmann, and P.~Mutzel.
\newblock Drawing planar partitions {II}: {HH-Drawings}.
\newblock In {\em WG'98}, volume 1517 of {\em LNCS}, 1998.

\bibitem{bs-pnpg-84}
J.~F. Buss and P.~W. Shor.
\newblock On the pagenumber of planar graphs.
\newblock In {\em STOC '84}, pages 98--100. ACM, 1984.

\bibitem{OGDF}
M.~Chimani, C.~Gutwenger, M.~J{\"u}nger, G.~Klau, K.~Klein, and P.~Mutzel.
\newblock {\em Handbook of Graph Drawing and Visualization: The Open Graph
  Drawing Framework}.
\newblock CRC-Press, 2012.

\bibitem{CorteseB05}
P.~F. Cortese and G.~{Di Battista}.
\newblock Clustered planarity.
\newblock In {\em SoCG '05}, pages 32--34, 2005.

\bibitem{dd-hgdvg-11}
G.~{Di Battista} and W.~Didimo.
\newblock {\em Handbook of Graph Drawing and Visualization: GDToolkit}.
\newblock CRC-Press, 2012.

\bibitem{dt-omtc-96}
G.~{Di Battista} and R.~Tamassia.
\newblock On-line maintenance of triconnected components with {SPQR}-trees.
\newblock {\em Algorithmica}, 15(4):302--318, 1996.

\bibitem{dt-opt-96}
G.~{Di Battista} and R.~Tamassia.
\newblock On-line planarity testing.
\newblock {\em SIAM J. Comput.}, 25:956--997, 1996.

\bibitem{fce-pcg-95}
Q.~Feng, R.~Cohen, and P.~Eades.
\newblock Planarity for clustered graphs.
\newblock In {\em ESA}, volume 979 of {\em LNCS}, pages 213--226, 1995.

\bibitem{ffr-opnupda-12}
F.~Frati, R.~Fulek, and A.~Ruiz-Vargas.
\newblock On the page number of upward planar directed acyclic graphs.
\newblock In {\em GD'12}, volume 7034 of {\em LNCS}, pages 391--402, 2012.

\bibitem{gt-upt-95}
A.~Garg and R.~Tamassia.
\newblock Upward planarity testing.
\newblock In {\em SIAM J. on Computing}, pages 436--441, 1995.

\bibitem{gm-lti-00}
C.~Gutwenger and P.~Mutzel.
\newblock A linear time implementation of {SPQR}-trees.
\newblock In {\em GD '00}, volume 1984 of {\em LNCS}, pages 77--90, 2001.

\bibitem{hn-tpbesg-09tr}
S.~Hong and H.~Nagamochi.
\newblock Two-page book embedding and clustered graph planarity.
\newblock TR [2009-004], Dept. of Applied Mathematics and Physics, University
  of Kyoto, Japan, 2009.

\bibitem{o-btvg-73}
L.~T. Ollmann.
\newblock On the book thicknesses of various graphs.
\newblock Cong. Num., VIII, page 459, 1973.

\bibitem{w-dcbe-02}
David~R. Wood.
\newblock Degree constrained book embeddings.
\newblock {\em J. of Algorithms}, 45(2):144--154, 2002.

\bibitem{y-epgfp-89}
M.~Yannakakis.
\newblock Embedding planar graphs in four pages.
\newblock {\em JCSS}, 38(1):36--67, 1989.

\end{thebibliography}

\end{document}